\overfullrule=0pt
\input harvmac
\def\a{{\alpha}}

\def\ad{{\dot a}}
\def\bd{{\dot b}}

\def\l{{\lambda}}
\def\lb{{\bar\lambda}}
\def\b{{\beta}}

\def\g{{\gamma}}
\def\G{{\Gamma}}

\def\d{{\delta}}

\def\e{{\epsilon}}
\def\s{{\sigma}}

\def\half{{1\over 2}}
\def\p{{\partial}}

\def\t{{\theta}}
\def\tb{{\bar\theta}}

\Title{\vbox{\hbox{~ }}}
{\vbox{
\centerline{\bf Ten-Dimensional Super-Twistors and Super-Yang-Mills}}}
\bigskip\centerline{Nathan Berkovits\foot{e-mail: nberkovi@ift.unesp.br}
}
\bigskip
\centerline{\it Instituto de F\'\i sica Te\'orica, UNESP-Universidade Estadual
Paulista}
\centerline{\it R. Dr. Bento T. Ferraz 271 - Bl. II, 01140-070, S\~ao Paulo, SP, Brasil}

\vskip .3in
Four-dimensional super-twistors provide a compact covariant
description of on-shell
${\cal N}=4$ d=4 super-Yang-Mills. In this paper, ten-dimensional super-twistors
are introduced which similarly provide a compact covariant description of
on-shell d=10 super-Yang-Mills. The super-twistor variables are 
$Z=(\lambda^\alpha,\mu_\alpha, \Gamma^m)$ where $\lambda^\alpha$ and
$\mu_\alpha$ are constrained bosonic d=10 spinors and $\Gamma^m$ is
a constrained fermionic d=10 vector. The Penrose map relates the
twistor superfield $\Phi(Z)$ with the d=10 super-Yang-Mills vertex operator
$\lambda^\alpha A_\alpha(x,\theta)$ which appears in the pure spinor formalism
of the superstring, and the cubic super-Yang-Mills amplitude is proportional
to the super-twistor integral
$\int dZ ~ \Phi_1 \Phi_2 \Phi_3$.

\Date{October 2009}

\newsec{Introduction}

In four dimensions, twistor variables were introduced in 1967 by Penrose as
an alternative description of spacetime in which light-like lines
replace points as the fundamental objects \ref\pen{R. Penrose,
``Twistor Algebra,'' J. Math. Phys. 8 (1967) 345.}. Instead of the usual
spacetime vector variable $x^\mu$ for $\mu=0$ to 3, Penrose's twistor
variables consist of bosonic two-component spinors $(\l^a,\mu_\ad)$
where $a,\ad=1$ to 2. The relation between these two descriptions
is given by 
\eqn\twistorrel{\mu_\ad = x_\mu\s^\mu_{a\ad}\l^a, \quad P^\mu = 
\l^a \s^\mu_{a\ad}
\lb^\ad}
where $P^\mu$ is the light-like momentum, $\s^\mu_{a\ad}$ are
the d=4 Pauli matrices, and $\lb^\ad$ is the canonical momentum variable
to $\mu_\ad$. These d=4 twistor variables transform linearly under
$SO(4,2)$ conformal transformations and provide a compact description
of massless states.

In 1978, Ferber \ref\ferber{A. Ferber, ``Supertwistors and conformal
supersymmetry,'' Nucl. Phys. B132 (1978) 55.}
generalized Penrose's twistors to four-dimensional super-twistors
consisting of the bosonic spinor variables $(\l^a,\mu_\ad)$ as well as
${\cal N}$ fermionic scalar variables $\eta^J$ for $J=1$ to ${\cal N}$
where ${\cal N}$ is the number of supersymmetries.
These super-twistor variables transform linearly under
superconformal transformations and are related to 
the usual $(x^\mu,\t^{a J},\tb^\ad_J)$ superspace variables by the map
\eqn\supertwistorrel{\mu_\ad  
= x_\mu\s^\mu_{a\ad}\l^a +\tb_{\ad J} \eta^J, \quad \eta^J = \l^a\t_a^J,}
$$P^\mu = \l^a \s^\mu_{a\ad} \lb^\ad, \quad q^a_{J} = \l^a \bar\eta_J, \quad
\bar q_\ad^J = \eta^J \lb_\ad$$
where $\lb^\ad$ is the canonical momentum variable to $\mu_\ad$ and
$\bar\eta_J$ is the canonical momentum variable to $\eta^J$.

When ${\cal N}=4$, these super-twistors provide a compact covariant
description of on-shell maximally supersymmetric $d=4$ super-Yang-Mills.
Expanding in powers of $\eta^J$, a scalar twistor superfield 
$\Phi(\l,\mu,\eta)$
of momentum $P^\mu = \l^a\s^\mu_{a\ad}\bar\pi^\ad$ (where $\bar\pi^\ad$
is the eigenvalue of the operator $\lb^\ad$)
has the expansion
\eqn\phifour{\Phi(\l,\mu,\eta) =e^{\mu_\ad\bar\pi^\ad}
( a_- +  \eta^J \bar s_J + \eta^J\eta^K \phi_{JK}+ 
(\eta^3)_J s^J + (\eta^4) a_+)}
where $(a_-, a_+)$ are the $(-1,+1)$ helicities of the gluon,
$(\bar s_J, s^J)$ are the $(-\half,+\half)$ helicities of the gluino,
and $\phi_{JK}$ are the six scalars. Recently, these super-twistors
have played an important role in simplifying the computation of d=4
super-Yang-Mills scattering amplitudes. Starting from the supersymmetric
expression for the MHV tree amplitude \ref\nair
{V. Nair, ``A current algebra for some gauge theory amplitudes,'' Phys.
Lett. B214 (1988) 215.}, Witten showed how to use super-twistors
to compute non-MHV super-Yang-Mills tree amplitudes \ref\witten
{E. Witten, ``Perturbative gauge theory as a string theory
in twistor space,'' Comm. Math. Phys. 252 (2004) 189, 
hep-th/0312171.}. These super-twistor
methods were further developed in hundreds of papers
and drastically simplify the
conventional Feynman diagram techniques for computing ${\cal N}=4$
d=4 super-Yang-Mills scattering amplitudes.

Despite this progress in computing ${\cal N}=4$ d=4 super-Yang-Mills
amplitudes using super-twistors, there has been very little discussion 
of super-twistors in higher dimensions. Although 
super-Yang-Mills is only conformally invariant in four dimensions, the most
natural formulation of maximally supersymmetric Yang-Mills is in
ten dimensions where the only physical fields are a gluon and
gluino. Moreover, super-Yang-Mills in ten dimensions is the low energy
limit of open superstring theory, and superstring theory and twistor theory
have many similar features \ref\wittwo{E. Witten,
``Twistor-Like Transform in Ten-Dimensions,''
Nucl. Phys. B266 (1986) 245.}
\ref\sorokin
{D. Sorokin, V. Tkach, D. Volkov and A.A. Zheltukhin, ``From 
the Superparticle Siegel Symmetry to the 
Spinning Particle Proper Time Supersymmetry,''
Phys. Lett. B216 (1989) 302.}. For example, both superstring theory and
twistor theory provide drastic simplifications to conventional 
Feynman diagram methods, at least when the external states are on-shell.
This suggests that any d=10 super-twistor description of super-Yang-Mills
might be related to superstring theory.

In this paper, a new d=10 super-twistor description will be introduced
which consists of the bosonic d=10 spinors $(\l^\a,\mu_\a)$ for $\a=1$
to 16, and the fermionic d=10 vector $\Gamma^m$ for $m=0$ to 9.
These super-twistor variables will be constrained to satisfy
\eqn\tencons{\l\g^m\l=0, \quad \mu_\a\l^\a = 0, 
\quad \mu\g^{mn}\l = 2 \G^m\G^n ,\quad
\G^m (\g_m\l)_\a =0}
where $\g^m_{\a\b}$ are the d=10 Pauli matrices satisfying
$\g^{(m}_{\a\b} \g^{n) \b\g} = 2\eta^{mn}\d_\a^\g$.
The first constraint of \tencons\ implies that $\l^\a$ is a d=10
pure spinor with 11 independent components, and the
remaining constraints of \tencons\ imply that $\mu_\a$ and $\G^m$
each have 5 independent components.
These d=10 super-twistor variables are related to the usual
d=10 superspace variables $(x^m,\t^\a)$ by the map
\eqn\tenmap{\mu_\a = x_m \g^m_{\a\b} \l^\b + \half\G^m (\g_m\t)_\a, \quad
\G^m = \l\g^m\t,}
$$P^m = \l\g^m\lb,\quad q_\a =  \bar\G_m (\g^m\l)_\a -
\G^m (\g_m\lb)_\a $$
where 
$\lb^\a$ is the canonical momentum variable to $\mu_\a$ and
$\bar\G_m$ is the canonical momentum variable to $\G^m$.

The use of bosonic pure spinor variables $(\l^\a,\mu_\a)$ to describe
higher-dimensional twistors has previously been discussed in \ref\hughston
{L.P. Hughston, ``The Wave Equation in Even Dimensions,''
in Further Advances in 
Twistor Theory, vol. 1, Research Notes in Mathematics 231, Longman, 26-27, 
1990\semi 
L.P. Hughston, ``A Remarkable Connection between the Wave Equation and Pure 
Spinors in Higher Dimensions,'' in Further Advances in Twistor Theory, 
vol. 1, Research Notes in Mathematics 231, Longman, 37-39, 1990.}
\ref\cherkis{N. Berkovits and S. Cherkis,
``Higher-dimensional twistor transforms using pure spinors,'' 
JHEP 0412 (2004) 049, hep-th/0409243.}
\ref\boels{R. Boels, ``Covariant representation theory 
of the Poincare algebra and some of its extensions,''
arXiv:0908.0738.},
and the d=10 super-twistor variables of \tencons\ can be understood as
a ``complexified'' version of the real d=10 super-twistor variables
introduced in \ref\btwist{N. Berkovits,
``A Supertwistor Description Of The 
Massless Superparticle In Ten-Dimensional Superspace,''
Nucl. Phys. B350 (1991) 193.}.
Unlike in four dimensions where the super-twistor
variables transform linearly under $d=4$
superconformal transformations,
the d=10 super-twistor variables transform linearly only under
d=10 super-Poincar\'e transformations.
Note that d=4 super-twistor variables involving fermionic vectors have
been discussed in \ref\zelh{A.A. Zheltukhin, ``Unification of twistors
and Ramond vectors,'' Phys. Lett. B658 (2007) 82, arXiv:0707.3453.},
and d=10 super-twistor variables involving fermionic
scalars have been discussed in 
\ref\uvarov{D. Uvarov, ``Canonical description of
D=10 superstring formulated in supertwistor space,'' J.Phys. A 42 (2009)
115202, arXiv:0804.0908.}.

To describe on-shell d=10 super-Yang-Mills, the twistor superfield
$\Phi(\l,\mu,\G)$ should satisfy the constraint $B\Phi=0$ where
\eqn\bone{B=(\l\g^m\lb) \bar\G_m - \half(\lb\g^m\lb) \G_m}
is a super-Poincar\'e
covariant operator.
The condition $B\Phi=0$ implies that $\Phi$ depends on only 4 of the
5 independent $\G$'s, so $\Phi$ has $2^4$ component fields as expected.
Expanding in $\G^m$, the component super-Yang-Mills
fields appear in the twistor superfield as
\eqn\expten{\Phi(\l,\mu,\G) =}
$$ e^{\mu_\a\bar\pi^\a}
(\bar s + \G_m a_-^m + \G_m\G_n s^{mn}
+ (\bar\pi\g_{mnpqr}\bar\pi) \G^m \G^n \G^p h^q a_+^r +
(\bar\pi\g_{mnpqr}\bar\pi) \G^m \G^n \G^p \G^q h^r s)$$
where $P^m = \l\g^m\bar\pi$, 
$h^m$ is any constant vector satisfying $h_m P^m =1$, the
d=10 gluon polarization 
has been split as $a^m = a^m_- + a^m_+$ with 
$a_-^m (\g_m\bar\pi)_\a = a_+^m (\g_m\l)_\a =0$,
and the
d=10 gluino polarization has been split as 
$\psi^\a = \bar\pi^\a \bar s + (\g_{mn}\l)^\a s^{mn} + \l^\a s$
with $s^{mn}(\g_n\bar\pi)_\a=0$.

Using the relation of \tenmap\
to map super-twistor variables into superspace
variables, one finds that $B$ maps into $\half u_\a P^m (\g_m D)^\a$
where $D_\a$ is the d=10 supersymmetric derivative and $u_\a$ is any
spinor satisfying $u_\a \l^\a=1$. And $\Phi$ of \expten\ maps
to $\l^\a A_\a(x,\t)$ where $A_\a(x,\t)$ is the super-Yang-Mills
spinor gauge superfield in the gauge $P^m (\g_m D)^\a A_\b=0$.

In the pure spinor formalism for the superstring
\ref\nbpure{N. Berkovits, ``Super-Poincar\'e
covariant quantization of the superstring,''
JHEP 0004 (2000) 018, hep-th/0001035.}, $V=\l^\a A_\a(x,\t)$
is the unintegrated open string vertex operator for super-Yang-Mills
and the composite $b$ ghost is $\half
u_\a P^m (\g_m D)^\a + ...$ where
$...$ involves non-minimal variables. $N$-point super-Yang-Mills
tree amplitudes are computed 
in this superstring formalism by evaluating the $\a'\to 0$ limit
of the disk correlation function
\eqn\diskcorr{{\cal A}_N = \langle V_1(z_1) V_2(z_2) V_3(z_3)
~\prod_{r=4}^N \int dz U_r(z_r) \rangle}
where $U_r(z_r) = b_{-1} V_r(z_r)$ is a dimension-one vertex operator
and $b_{-1}$ denotes the single pole with $b$.

Since $\Phi$ maps to $V$ and $B$ maps to the $b$ ghost, it is natural
to try to formulate a similar prescription for d=10 super-Yang-Mills tree
amplitudes in terms of twistor superfields. As in d=4, it 
will be shown that the cubic super-Yang-Mills amplitude in d=10
is proportional to the super-twistor integral
\eqn\treetw{ \int d^{10}\l d^5\mu d^5\G~\Phi_1 \Phi_2 \Phi_3.}
However, it is not yet understood how to obtain the correct proportionality
factor coming from momentum conservation for this cubic d=10 amplitude. 
Furthermore,
for higher-point amplitudes, the appropriate super-twistor
prescription is not known and will require the construction of a worldsheet
action for the super-twistor variables.

In section 2 of this paper, the four-dimensional super-twistor description
of ${\cal N}=4$ d=4 super-Yang-Mills is reviewed. In section 3, the
d=10 super-twistor variables are introduced, the twistor superfield
for d=10 super-Yang-Mills is constructed, and the super-twistor prescription
for super-Yang-Mills tree amplitudes is discussed.

\newsec{Review of Four-Dimensional Super-Twistors}

\subsec{${\cal N}=4$ d=4 super-twistor variables}

${\cal N}=4$ d=4 super-twistor variables consist of the
bosonic Weyl and anti-Weyl spinors $(\l^a,\mu_\ad)$ for $a,\ad=1$ to 2
and the fermionic scalars $\eta^J$ for $J=1$ to 4. Note that unlike
the usual superspace variables $(x^\mu,\t^{aJ},\tb^\ad_J)$ for $\mu=0$
to 3, super-twistor variables carry the opposite statistics from those
expected by the spin-statistics relation. 

These variables transform
linearly under $PSU(2,2|4)$ superconformal transformations which
are generated by
\eqn\sucon{P^\mu = \l^a \s^\mu_{a\ad}\lb^\ad, \quad q_{aJ}=\l_a\bar\eta_J,\quad
\bar q_\ad^J =  \eta^J \lb_\ad, \quad
M^{\mu\nu} =\half (\s^{\mu\nu})_a^b \l^a \bar\mu_b +\half (\s^{\mu\nu})_\ad^\bd
\mu_\bd \lb^\ad,}
$$
R_K^J = \eta^J\bar\eta_K, \quad
K^\mu = \s^\mu_{a\ad} \mu^\ad\bar\mu^a,\quad D = \l^a\bar\mu_a -
\mu_\ad\bar\l^\ad,\quad s_a^J = \eta^J \bar\mu_a, \quad
\bar s_{\ad J} = \mu_{\ad}\bar\eta_J, $$
where $(\bar\mu_a,\bar\l^\ad,\bar\eta_J)$ are the canonical momenta
variables to $(\l^a,\mu_\ad, \eta^J)$ and $\s^\mu_{a\ad}$ are the
d=4 Pauli matrices. These superconformal generators all commute with the
generator
\eqn\proj{H = \l^a\bar\mu_a + \mu_\ad \bar\l^\ad + \eta^J\bar\eta_J}
which defines the ``projective weight''.

The above super-twistor variables are related to the usual
${\cal N}=4$ d=4 superspace variables $(x^\mu,\t^{aJ},\tb^\ad_J)$
by the relation
\eqn\relfour{\mu_\ad = x_\mu\s^\mu_{a\ad}\l^a + \tb_{\ad J}\eta^J,\quad
\eta^J = \l^a \t_a^J.}
One can easily check that this map is invariant under the above
$PSU(2,2|4)$ superconformal transformations if one defines $(x^\mu,\t^{aJ},
\tb^\ad_J)$ to transform in the standard manner. 

\subsec{Twistor superfield for d=4 super-Yang-Mills}

Just as on-shell ${\cal N}=4$ d=4 super-Yang-Mills can be
described using gauge and field-strength
superfields depending on $(x,\t,\tb)$ superspace
variables, it can also be described by a scalar twistor superfield
$\Phi(\l,\mu,\eta)$. The map of \relfour\
relates this twistor superfield with the spacetime field-strength superfield
$F_{\ad\bd}(x,\t,\tb)$ whose $\t=\tb=0$ component is the
linearized self-dual field-strength $f_{\ad\bd} = (\s^{\mu\nu})_{\ad\bd}
\p_\mu a_\nu$.

To show this relation, first note that if $F_{\ad\bd}$ is written
in momentum space where the light-like momentum satisfies $P^\mu = 
\l^a\s^\mu_{a\ad}\bar\pi^\ad$, $F_{\ad\bd}$ can be expressed as
$F_{\ad\bd} = \bar\pi_\ad\bar\pi_\bd F(x,\t,\tb)$
where the $\t=\tb=0$ component of $F$ is the $-1$ helicity component
of the gluon. In other words, if the gluon polarization $a^\mu$ is
split as 
\eqn\gluonsplit{a^\mu = (\e^a\s^\mu_{a\ad}\bar\pi^\ad) a_- +
(\bar\e^\ad\s^\mu_{a\ad}\l^a) a_+} 
where $(\e^a,\bar\e^\ad)$ are arbitrary spinors satisfying
$\e^a\l_a = \bar\e^\ad\bar\pi_\ad=1$,
$a_-$ is the $\t=\tb=0$ component of $F$.
Furthermore, the superspace constraints $\bar D_{\dot c}^J F_{\ad\bd} =0$
and $\s^\mu_{c{\dot c}} \p_\mu D^c_J F_{\ad\bd}=0$ imply that 
$F$ satisfies
\eqn\Fcons{\bar D_\ad^J F = \l_a D_J^a F=0} 
where $D_{aJ}= {\p\over{\p\t^{aJ}}}-\half
(\s^\mu\tb)_{aJ} \p_\mu$ and $\bar D_\ad^J = 
{\p\over{\p\tb^\ad_J}}-\half
(\s^\mu\t)^J_{\ad} \p_\mu$
are the ${\cal N}=4$ d=4 supersymmetric derivatives.

To relate $F(x,\t,\tb)$ to $\Phi(\l,\mu,\eta)$, define the scalar twistor
superfield with momentum $P^\mu = \l^a\s^\mu_{a\ad}\bar\pi^\ad$
as
\eqn\twistorform{\Phi(\l,\mu,\eta) = 
e^{\mu_\ad \bar\pi^\ad} f(\eta^J)}
where 
$f(\eta^J)$ is an arbitrary function of $\eta^J$. 
The map relating $F(x,\t,\tb)$ and $\Phi(\l,\mu,\eta)$ 
is defined by 
\eqn\penrose{F(x,\t,\tb)= \widetilde\Phi
(\l,x,\t,\tb)}
where $F(x,\t,\tb)$ has momentum $P^\mu=\l\s^\mu\bar\pi$ and
$\widetilde\Phi(\l,x,\t,\tb)$ is obtained from $\Phi(\l,\mu,\eta)$
by setting 
$\mu_\ad = x_\mu\s^\mu_{a\ad}\l^a + \tb_{\ad J}\eta^J$ and
$\eta^J = \l^a \t_a^J$ as in \relfour. It is easy to use 
\relfour\ to verify that
$\bar D_\ad^J \widetilde\Phi = \l^a D_{aJ}\widetilde\Phi=0$, so
\Fcons\ is satisfied.
Using the identification of \penrose\ with $F(x,\t,\tb)$, one
learns that 
\eqn\flearn{f(\eta^J) = a_- + \eta^J \bar s_J +
\eta^J\eta^K \phi_{JK} + (\eta)^3_J s^J + (\eta)^4 a_+}
where $a_\pm$ are defined in \gluonsplit, $s^J$ and $\bar s_J$ 
are the gluinos of $\pm \half$ helicity defined by
$\psi_a^J = \pi_a s^J$ and $\bar\psi_{\ad J} = \bar\pi_\ad \bar s_J$,
and $\phi_{JK}$ are the six scalars.
Note that $f(\eta^J)$ has $+2$ projective weight since under
$\l^a\to c\l^a$ and $\bar\pi^\ad\to c^{-1}\bar\pi^\ad$,
\eqn\projfour{(a_-, \bar s_J, \phi_{JK}, s^J, a_+)\to
(c^2 a_-, c \bar s_J, \phi_{JK}, c^{-1} s^J, c^{-2} a_+).}

\subsec{d=4 super-Yang-Mills tree amplitudes}

Four-dimensional super-twistors have recently been used to
compute super-Yang-Mills tree amplitudes where
tree amplitudes of different helicity
violation involve curves in twistor space of different degree
\witten\ref\roiban
{R. Roiban, M. Spradlin and A. Volovich,  
''On the tree level S matrix of Yang-Mills theory,''
Phys. Rev. D70 (2004) 026009, 
hep-th/0403190.}. Only the
degree zero curve where $\l^a$ is constant
will be discussed here, which is non-vanishing
for cubic ``self-dual'' amplitudes, i.e. the supersymmetric
completion of amplitudes involving
two gluons of $-1$ helicity and one gluon of $+1$ helicity. Note that
although the cubic amplitude vanishes for real momentum in signature
$(d-1,1)$, it is non-vanishing 
for real momentum in signature $({d\over 2}, {d\over 2})$.

In signature $(2,2)$, there are two possible ways for the momentum
conservation condition $\sum_{r=1}^3 P_{(r)}^\mu = 
\sum_{r=1}^3 (\l_{(r)} \s^\mu \bar\pi_{(r)}) =0$ to be satisfied.
Either $\l_{(1)}^a=\l_{(2)}^a=\l_{(3)}^a$ and $\sum_{r=1}^3\bar\pi_{(r)}^\ad=0$,
or
$\bar\pi_{(1)}^\ad=\bar\pi_{(2)}^\ad=\bar\pi_{(3)}^\ad$ and $\sum_{r=1}^3
\l_{(r)}^a=0$. The first solution corresponds to the ``self-dual'' amplitude
involving a degree zero curve where $\l^a$ is constant, whereas  
the second solution corresponds to the ``anti-self-dual'' amplitude involving
a degree one curve where $\l^a$ is non-constant.

Suppose one uses projective invariance to scale $\l^1_{(1)} = \l^1_{(2)} =
\l^1_{(3)}=1$. Then if one defines
\eqn\phir{\Phi_{(r)}(\l,\mu, \eta) = \d(\l^2-\l^2_{(r)}) 
e^{\mu_\ad\bar\pi^\ad_{(r)}} f_{(r)}(\eta)}
where $f_{(r)}(\eta)$ is defined in \flearn, 
the cubic self-dual super-Yang-Mills amplitude can be expressed
as the super-twistor integral
\eqn\cubicsd{{\cal A} =\int d\l^2 \int d^2\mu \int d^4\eta 
Tr([\Phi_{(1)},\Phi_{(2)}]\Phi_{(3)})}
where the trace is over the color indices of $\Phi$.
The integral over $\int d\l^2 \int d^2\mu \int d^4\eta$ is easily 
performed and gives
\eqn\ampcub{
{\cal A} =  
\d(\l_{(3)}^2-\l_{(1)}^2)\d(\l_{(3)}^2-\l_{(2)}^2)\d^2 (\sum_r \bar\pi_{(r)}) 
Tr([a^{(1)}_- ,a^{(2)}_-] a^{(3)}_+ + ...)}
$$=(\bar\pi_{(1)}^\ad\bar\pi_{(2)\ad})
 \d^2(\sum_r \l_{(r)}^2 \bar\pi_{(r)})
 \d^2(\sum_r \bar\pi_{(r)}) 
Tr([a^{(1)}_- ,a^{(2)}_-] a^{(3)}_+ + ...)$$
$$=(\bar\pi_{(1)}^\ad\bar\pi_{(2)\ad})
 \d^4(\sum_r P_{(r)})
Tr([a^{(1)}_- ,a^{(2)}_-] a^{(3)}_+ + ...),$$
which is the correct expression for the self-dual super-Yang-Mills
amplitude where $...$ is the supersymmetric completion of the self-dual
gluon amplitude.

\newsec{Ten-Dimensional Super-Twistors}

\subsec{$d=10$ super-twistor variables}

As discussed in \hughston\cherkis\boels, 
the natural generalization of four-dimensional
twistor variables $(\l^a,\mu_\ad)$ to higher dimensions is 
$(\l^\a,\mu_\a)$ where $\l^\a$ is a pure spinor and $\mu_\a$ is
related to the spacetime variables $x^m$ by the map $\mu_\a = 
x^m (\g_m\l)_\a$. In ten dimensions, a Weyl spinor has
16 components (i.e. $\a=1$ to 16),
and a pure spinor must satisfy
$\l\g^m\l=0$ which implies that $\l^\a$ has only 11 independent
components. Furthermore, $\mu_\a = x^m (\g_m\l)_\a$ implies
that $\mu_\a$ satisfies $\mu_\a\l^\a = \mu\g^{mn}\l =0$, which implies
that $\mu_\a$ has only 5 independent components. Note that in spacetime with
signature $(9,1)$, $(\l^\a,\mu_\a)$ are complex variables. But just as
four-dimensional twistors are real variables in signature $(2,2)$,
ten-dimensional twistors are real variables in signature $(5,5)$.
In this paper, we shall choose the signature $(5,5)$ so that $\l^\a$
and $\lb^\a$ are independent real variables.

As in four dimensions, $\mu_\a$ can be interpreted as the canonical
momentum variable to $\bar\l^\a$ where the light-like spacetime momentum
is $P^m = \l^\a \g^m_{\a\b}\lb^\b$. Note that $\lb^\a$ is not required
to be a pure spinor, and $P^m P_m=0$ follows from the d=10 gamma-matrix
identity $\g_{m\a(\b}\g^m_{\g\d)}=0$ together with $\l\g^m\l=0$.

To generalize to ten-dimensional super-twistors, one introduces a
fermionic vector $\G^m$ which is constrained to satisfy
$\G^m (\g_m\l)_\a=0$. So $\G^m$ has 5 independent components.
As in four dimensions, the statistics of the
fermonic twistor variable is opposite from the statistics one would expect
from its Lorentz representation.
One also modifies the constraints on $\mu_\a$ to 
$\mu_\a\l^\a = \mu\g^{mn}\l - 2\G^m\G^n=0$. 

So the ten-dimensional super-twistor space is defined by the variables
$(\l^\a,\mu_\a,\G^m)$ which are constrained to satisfy
\eqn\tenc{\l\g^m\l=0,\quad \mu_\a\l^\a=0,\quad
\mu\g^{mn}\l - 2\G^m\G^n =0,\quad \G^m (\g_m\l)_\a =0.}
These constraints imply that $\mu_\a$ and $\G^m$ can be expressed 
in terms of d=10 superspace variables $(x^m,\t^\a)$ as
\eqn\tenrel{\mu_\a = x_m (\g^m\l)_\a + \half\G^m(\g_m\t)_\a, \quad 
\G^m = \l\g^m\t,}
which closely resembles the four-dimensional relation of \relfour.
Furthermore, these super-twistor variables transform linearly
under d=10 super-Poincar\'e transformations which are
generated by
\eqn\superpten{P^m = \l\g^m\lb,\quad q_\a = 
(\g^m\l)_\a \bar\G_m -\G^m (\g_m\lb)_\a,\quad 
M^{mn} = \half\l\g^{mn}\bar\mu + \half\mu\g^{mn}\lb + \G^{[m}\bar\G^{n]},}
where $(\bar\mu_\a,\lb^\a, \bar\G_m)$ are canonical momentum variables
for $(\l^\a,\mu_\a, \G^m)$. 

It is easy to verify that the generators of
\superpten\ commute with the constraints of \tenc\ and form a d=10
super-Poincar\'e algebra. As in four dimensions, the operator
\eqn\projten{H = \l^\a\bar\mu_\a + \mu_\a\lb^\a + \G^m\bar\G_m}
defines the projective weight and commutes with the super-Poincar\'e
generators. Finally, it will be useful to define the fermionic operator
\eqn\Bg{B = (\l\g^m\lb)\bar \G_m - \half \G^m (\lb\g_m\lb)}
which commutes with both the constraints of \tenc\ and with the
super-Poincar\'e generators of \superpten.

If one sets the fermionic variables $\G^m$ and $\bar\G_m$ to zero,
the d=10 Poincar\'e algebra can be extended to a conformal algebra
by including the generators $K^m = \mu\g^m\bar\mu$ and
$D = \mu\lb - \l\bar\mu$. However, after including $\G^m$ and $\bar\G_m$,
there is no obvious way to extend the d=10 super-Poincar\'e algebra
to a superconformal algebra. This is of course not surprising since
d=10 super-Yang-Mills is not superconformally invariant.

\subsec{d=10 twistor superfield}

In this section, it will be shown that on-shell d=10 super-Yang-Mills
is described by a scalar twistor superfield $\Phi(\l,\mu,\G)$
of $+1$ projective weight which is annihilated by the $B$ operator of
\Bg. This twistor superfield can be mapped to the spacetime superfield
$V=\l^\a 
A_\a(x,\t)$ which appears in the pure spinor formalism, and
the condition of $+1$ projective weight is related to the
$+1$ ghost-number of $V$.
The condition that $B\Phi=0$ comes from a gauge-fixing condition
on $A_\a$ and implies that
$\Phi$ depends on only 4 of the 5 $\G$'s, which is the same number
of fermionic variables as in the four-dimensional super-twistor. 

In the pure spinor formalism for the superparticle or superstring,
linearized on-shell d=10 super-Yang-Mills is described by the
vertex operator $V=\l^\a A_\a(x,\t)$ satisfying $QV=0$ where
$Q=\l^\a D_\a$, $D_\a = {\p\over{\p\t^\a}} -\half 
(\g^m\t)_\a {\p\over{\p x^m}}$,
$\l^\a$ is a d=10 pure spinor, $A_\a(x,\t)$ is the spinor gauge superfield
satisfying $D_{(\a}A_{\b)} = \g^m_{\a\b} A_m$, and $A_m(x,\t)$
is the vector gauge superfield \ref\howe{P.S. Howe,
``Pure spinors lines in superspace and ten-dimensional supersymmetric 
theories.''
Phys. Lett. B258 (1991) 141.}. The gauge superfields $A_\a$ and $A_m$
are defined up to the linearized gauge transformations $\d A_\a = D_\a\Omega$
and $\d A_m = \p_m \Omega$.

A convenient gauge-fixing condition for $A_\a$
is $\p_m (\g^m D)_\a A_\b(x,\t)=0$. This gauge-fixing condition implies
that $\p^m A_m=0$ and can be solved in a plane-wave basis with momentum $P_m$
by
\eqn\gaugesol{A_\a = h^m (\g_m W)_\a, \quad A_n = h^m F_{mn}}
where $h^m$ is any constant vector satisfying $h^m P_m=1$,
$W^\a ={1\over{10}} \g^{m\a\b} (D_\b A_m -\p_m A_\b)$ is the superfield-strength
whose $\t=0$ component is the gluino, and $F_{mn} = \p_{[m} A_{n]}$
is the superfield-strength whose $\t=0$ component is the gluon field strength.

To relate $V=\l^\a A_\a$ with a twistor superfield, suppose that the
momentum $P^m$ satisfies $P^m = \l^\a\g^m_{\a\b}\bar\pi^\b$ for some
$\bar\pi^\b$. Then the superfield identity
$D_\a W^\b =-{1\over 4}
 (\g^{mn})_\a{}^\b F_{mn}$ implies that $V = (\l \g^m W) h_m$ 
satisfies
\eqn\Vcons{(\l^\a D_\a)V = (\l\g^{mn}D) V = (\bar\pi^\a D_\a) V =0.}

Since the momentum $P^m = \l\g^m\bar\pi$ is invariant under the transformation
$\d\bar\pi^\a = (\g^{mn}\l)^\a \Omega_{mn}$ for arbitrary $\Omega_{mn}$,
one can choose $\bar\pi^\a$ 
so that it is a pure spinor satisfying $\bar\pi\g^m\bar\pi=0$.
The d=10
twistor superfield will then be defined in analogy with \twistorform\  as 
\eqn\tentw{\Phi (\l,\mu,\G) = e^{\mu_\a \bar\pi^\a} f(\G^m).}
To satisfy $B\Phi=0$ where $B$ is defined in \Bg, $f(\G^m)$
must satisfy $(\l\g^m\bar\pi) {\p\over{\p\G^m}} f=0$ 
which implies that $f(\G^m)$
depends on only four of the five independent $\G$'s.

The map relating $V=\l^\a A_\a$ with $\Phi(\l,\mu,\G)$ will be defined as
in \penrose\ by
\eqn\tenpenrose{V(\l, x,\t) = \widetilde\Phi
(\l,x,\t)}
where $V(\l,x,\t)$ has momentum $P^m = \l\g^m\bar\pi$ and
$\widetilde\Phi(\l,x,\t)$ is obtained from $\Phi(\l,\mu,\G)$
by setting 
$\mu_\a = x_m (\g^m \l)_\a + \half\G^m (\g_m\t)_\a$ and
$\G^m = \l\g^m \t$ as in \tenrel. It is easy to use 
\tenrel\ to verify that
$(\l^\a D_\a) \widetilde\Phi = (\l\g^{mn} D)\widetilde\Phi=(\bar\pi^\a D_\a)
\widetilde \Phi = 0$, so
\Vcons\ is satisfied.
Using the identification of \tenpenrose\ with $V(\l, x,\t)$, one
learns (up to constant coefficients) that 
\eqn\tenflearn{
f(\G^m) = \bar s + \G_m a_-^m + \G_m\G_n s^{mn}
+ (\bar\pi\g_{mnpqr}\bar\pi) \G^m \G^n \G^p h^q a_+^r +
(\bar\pi\g_{mnpqr}\bar\pi) \G^m \G^n \G^p \G^q h^r s}
where $P^m = \l\g^m\bar\pi$, 
$h^m$ is any constant vector satisfying $h_m P^m =1$, the
d=10 gluon polarization 
has been split as $a^m = a^m_- + a^m_+$ with 
$a_-^m (\g_m\bar\pi)_\a = a_+^m (\g_m\l)_\a =0$,
and the
d=10 gluino polarization has been split as 
$\psi^\a = \bar\pi^\a \bar s + (\g_{mn}\l)^\a s^{mn} + \l^\a s$ 
with $s^{mn}(\g_n\bar\pi)_\a=0$.
Note that $f(\G^m)$ is annihilated by  $P^m {\p\over{\p\G^m}}$
and has $+1$ projective weight since under
$\l^a\to c\l^a$ and $\bar\pi^\a\to c^{-1}\bar\pi^\a$,
\eqn\projten{(a_-^m, a_+^m) \to (a_-^m, a_+^m) \quad{\rm and} \quad
(\bar s, s^{mn}, s) \to
(c \bar s, c^{-1} s^{mn}, c^{-1} s).}
Furthermore, $\bar\pi\g^m\bar\pi=0$ implies that $f(\G^m)$ is
independent of the explicit choice of $h^m$.

It might seem surprising that unlike the ${\cal N}=4$ d=4 twistor
superfield which is bosonic, the d=10 twistor superfield of \tentw\
is fermionic.
This is related to the fact that upon dimensional reduction to d=4,
$\G^m = \l\g^m\t$ involves the chiral d=4 $\t_a^J$'s for $J=1$
to 3 in the linear
combinations $\eta^J = \l^a \t_a^J$, 
but also involves
the antichiral d=4 $\bar\t^\ad_4$'s in the combinations
$\l\s^\mu \tb_4$. So the dimensional reduction
of the d=10 twistor superfield is not the d=4 twistor superfield 
of \twistorform.

\subsec{d=10 super-Yang-Mills tree amplitudes}

Since the d=10 super-twistors closely resemble the ${\cal N}=4$ d=4
super-twistors, it is natural to try to generalize the super-twistor
prescription for computing ${\cal N}=4$ d=4 super-Yang-Mills tree
amplitudes to ten dimensions. Although cubic super-Yang-Mills amplitudes
vanish for real momenta in spacetime signature $(9,1)$, they are
non-vanishing in signature $(5,5)$ where pure spinors have 11 real
components. To analyze the kinematics in this signature, it is convenient
to break manifest $SO(5,5)$ Lorentz invariance to an $SL(5)$ subgroup
such that a spinor $\l^\a$ decomposes into $(1,10,5)$ representations
which will be denoted as $(\l^+,\l_{jk},\l^j)$ for $j=1$ to 5. 
When $\l^\a$ is a pure spinor, the constraint $\l\g^m\l=0$ can be solved
by setting $\l^j = {1\over 8} (\l^+)^{-1} \e^{jklmn}\l_{kl}\l_{mn}$,
where it is assumed that the component $\l^+$ is non-zero.

If the momenta of the three external states are 
$P^m_{(r)} = \l_{(r)}\g^m \bar\pi_{(r)}$, one can use the invariance
$\d\bar\pi_{(r)}^\a = (\g^{mn}\l_{(r)})^\a \Omega_{(r)mn}$ 
(and the condition that
$\l^+_{(r)}\neq 0$) to fix $\bar\pi_{(r)}^+ = \bar\pi_{(r)jk}=0$.
So the only non-zero components of $\bar\pi_{(r)}^\a$ 
are in the 5 representation,
and the ten components of $P^m_{(r)}$ decompose under $SL(5)$ into
\eqn\decom{P^j_{(r)} = \l^+_{(r)}\bar\pi^j_{(r)}, \quad
P_{(r)j} = \l_{(r)jk}\bar\pi^k_{(r)}.}
Now if one uses projective invariance to scale $\l^+_{(1)}=\l^+_{(2)}
=\l^+_{(3)} =1$, momentum conservation implies as in d=4 that 
$\sum_{r=1}^3 \bar\pi_{(r)}^\a=0$. However, unlike in d=4, momentum
conservation does not imply that $\l_{(1)}^\a= \l_{(2)}^\a= \l_{(3)}^\a$,
and this will lead to a missing proportionality factor in the d=10
super-twistor prescription.

The natural d=10 generalization of the d=4 super-twistor prescription of 
\cubicsd\ is 
\eqn\tengen{{\cal A} = \int d^{10}\l \int d^5\mu \int d^5 \G ~
Tr ([\Phi_{(1)},\Phi_{(2)}]\Phi_{(3)})}
where $\Phi_{(r)}$ is defined as 
\eqn\tenphir{\Phi_{(r)}(\l,\mu,\G) = \d^{10}(\l_{jk} - \l_{(r)jk})
e^{\mu_j \bar\pi^j_{(r)}} f_{(r)}(\G)}
and $f_{(r)}(\G)$ is defined in \tenflearn. The integral 
$\int d^5 \G$ will be defined as
\eqn\intgg{\int d^5 \G ~F(\G) = {1\over{5!}}(\l\g^{mnpqr}\l) 
{\p\over{\p\G^m}}
{\p\over{\p\G^n}}
{\p\over{\p\G^p}}
{\p\over{\p\G^q}}
{\p\over{\p\G^r}} ~F(\G),}
which is consistent with
the constraint $\G^m (\g_m\l)_\a =0$ since
$(\l\g^{mnpqr}\l) 
{\p\over{\p\G^m}}~\G^s(\g_s\l)_\a=0$.

Performing the integration 
$\int d^{10}\l \int d^5\mu \int d^5 \G $, one finds that 
\eqn\tensol{{\cal A} = \d^{10}(\l_{(3)jk}-\l_{(1)jk})
\d^{10}(\l_{(3)jk}-\l_{(2)jk}) \d^5(\sum_{r=1}^3 \bar\pi_{(r)}^j)}
$$Tr ([a_{(1)}^m, a_{(2)}^n] P_{(3)m} a_{(3)n} + \psi_{(1)}\g^m\psi_{(2)}
a_{(3)m} + {\rm ~~permutations~ of~ (1,2,3)~~}).$$
This would be the correct cubic super-Yang-Mills amplitude if
$$\d^{10}(\l_{(3)jk}-\l_{(1)jk})
\d^{10}(\l_{(3)jk}-\l_{(2)jk}) \d^5(\sum_{r=1}^3 \bar\pi_{(r)}^j)$$
were equal to 
\eqn\wereeq{\d^{10}(\sum_{r=1}^3 P^m_{(r)})= 
\d^5 (\sum_{r=1}^3 \l_{(r)jk} \bar\pi^k_{(r)}) 
\d^5(\sum_{r=1}^3 \bar\pi^j_{(r)}).}
However, as remarked earlier, $\sum_r P_{(r)}^m=0$ does not imply
$\l_{(1)jk}= \l_{(2)jk}= \l_{(3)jk}$, so the first line of \tensol\
is too restrictive.

Despite this incorrect proportionality factor, it is remarkable that
$$\int d^5 \G ~f_{(1)}(\G) f_{(2)}(\G) f_{(3)}(\G)$$
correctly reproduces
the polarization dependence of the cubic super-Yang-Mills amplitude in
the second line of \tensol. This is related to the fact that 
$f_{(r)}(\G)$ is mapped to $V_{(r)} = \l^\a A_{(r)\a}(x,\t)$, and the
cubic super-Yang-Mills amplitude prescription in the pure spinor
formalism is ${\cal A} = \langle V_{(1)} V_{(2)} V_{(3)} \rangle$
where the normalization of $\langle~\rangle$ is defined by
\eqn\measure{ \langle (\l\g^m\t)(\l\g^n\t)(\l\g^p\t)(\t\g_{mnp}\t)\rangle=1.}
Note that $\G^m = \l\g^m\t$ implies that
$\G^m\G^n\G^p\G^q\G^r$ 
is proportional to
\eqn\propgam{(\l\g^{mnpqr}\l)(\l\g^s\t)(\l\g^t\t)(\l\g^u\t)(\t\g_{stu}\t).} 
So \intgg\ implies that
$\int d^5 \G~ F(\G)$ is proportional to $\langle \widetilde F(\l,\t)\rangle$
where $\widetilde F(\l,\t)$ is obtained from $F(\G)$ by setting
$\G^m = \l\g^m\t$.

Because of its close relationship with the pure spinor formalism,
it might be possible to get intuition about a super-twistor
prescription for $N$-point tree amplitudes from the pure spinor
formalism. The $N$-point tree amplitude prescription using this
superstring formalism is 
\eqn\purep{{\cal A} = \langle V_{(1)}(z_1) V_{(2)}(z_2) V_{(3)}(z_3)
\prod_{r=4}^N \int dz_r U_{(r)}(z_r)\rangle}
where $V=\l^\a A_\a$ is the dimension zero unintegrated vertex operator,
$U = \p\t^\a A_\a + (\p x^m -\half \t\g^m\p\t) A_m + ...$
is the dimension one vertex operator satisfying $QU = \p V$, and
$Q= \int dz \l^\a d_\a$ is the BRST operator.

In this formalism, the composite operator
\eqn\bgh{ b = \half(\p x^m -\half\t\g^m\p\t)(u\g_m d) + ...} 
satisfies $\{Q,b\}=T$ where $u_\a\l^\a =1$, $T$ is the stress tensor, $d_\a$
is the worldsheet variable corresponding to $D_\a$,
and $...$ involves non-minimal variables. After including dependence
on the non-minimal variables, one can choose Siegel gauge for $V$ 
\ref\aisaka{N. Berkovits
and Y. Aisaka, ``Pure Spinor Vertex Operators in Siegel Gauge and Loop 
Amplitude Regularization.''
JHEP 0907 (2009) 062, 
arXiv:0903.3443.}
in which
$b_0 V=0$ where $b_n V$ denotes the pole of order $(2+n)$ in the OPE of
$b$ and $V$. In this gauge, the integrated vertex operator can be expressed
as $\int dz U = \int dz b_{-1} V$ and the tree amplitude prescription of 
\purep\
can be expressed in terms of $V$ and $b$ as 
\eqn\purept{{\cal A} = \langle V_{(1)}(z_1) V_{(2)}(z_2) V_{(3)}(z_3)
\prod_{r=4}^N \int dz_r b_{-1} V_{(r)}(z_r)\rangle.}

If $d_\a$ is defined as $d_\a = q_\a - P^m (\g_m\t)_\a$ where
$q_\a = (\g^m\l)_\a \bar\G_m -
\G^m (\g_m\bar\l)_\a$ as in \superpten,
one finds that 
\eqn\Bfind{\half P_m (u\g^m d) = \half
P_m (u\g^m q)
=  (u_\a \l^\a) [P^m \bar\G_m - \half \G_m (\lb\g^m\lb)]
= B}
where the identity 
$P_m (\g^m\lb)_\a = -\half (\lb\g^m\lb)(\g_m\l)_\a$ has been used.
So the super-Poincar\'e invariant operator $B$ of \Bg\ is related to the
composite $b$ ghost. This suggests defining the $N$-point tree amplitude
super-twistor prescription as 
\eqn\pureptwis{{\cal A} = \int d^{10}\l\int d^5\mu \int d^5\G~
 \Phi_{(1)}(z_1) \Phi_{(2)}(z_2) \Phi_{(3)}(z_3)
\prod_{r=4}^N \int dz_r B_{-1} \Phi_{(r)}(z_r).}
However, in order to evaluate this amplitude prescription, one first
needs to define a worldsheet action for the super-twistor variables
and compute their OPE's. If this can be done, it seems reasonable to
conjecture that the resulting twistor string theory will be
related to the usual d=10 superstring theory by a field redefinition.
It is intriguing that the bosonic super-twistor variables $\l^\a$
resemble the the bosonic ghost variables in the pure spinor formalism, 
whereas the fermionic super-twistor variables $\G^m$ resemble the fermionic
matter variables in the RNS formalism. In fact, a recent proposal in 
\ref\relb{N. Berkovits, ``Explaining the 
Pure Spinor Formalism for the Superstring,''
JHEP 0801 (2008) 065,
arXiv:0712.0324,}
for relating the pure spinor and RNS worldsheet variables defined
fermionic variables $\G^m = \l\g^m\t$ and $\bar\G_m = u \g_m d$ which
were related to the RNS variables by twisting \ref\baulieu{L. Baulieu,
``Transmutation of pure 2-D supergravity into topological 2-D gravity and 
other conformal theories,'' Phys. Lett. B288 (1992) 59, hep-th/9206019.} as
\eqn\rnsrel{\psi^m = \g^{-1} \G^m + \g \bar\G^m}
where $\psi^m$ is the RNS fermionic matter variable and
$\g$ is the RNS bosonic ghost variable.
\vskip 15pt

{\bf Acknowledgements:}
I would
like to thank Zvi Bern, Sergei Cherkis, Radu Roiban and Anthony Zee
for useful discussions, and
CNPq grant 300256/94-9
and FAPESP grant 09/50639-2 for partial financial support,

\listrefs

\end